\documentclass{IOS-Book-Article}

\usepackage{mathptmx}
\usepackage{amsmath}

\DeclareMathAlphabet{\mathcal}{OMS}{cmsy}{m}{n}

\usepackage{amssymb}  % Provides \mathbb{} for math blackboard bold letters
\usepackage{soul}\setuldepth{article}
\def\hb{\hbox to 11.5 cm{}}

\begin{document}

\pagestyle{headings}
\def\thepage{}
\begin{frontmatter}              % The preamble begins here.

\title{Norms, Values, and other Ethical Constructs: Towards a Unifying Architecture for Technical Governance}

\markboth{}{October 2026\hb}

\author[A]{\fnms{Giovanni} \snm{Sileno}%\orcid{....-....-....-....}%
\thanks{Corresponding Author: Giovanni Sileno, \url{g.sileno@uva.nl}}} and
\author[B]{\fnms{Tomasz} \snm{Zurek}}%\orcid{....-....-....-....}}

\runningauthor{Sileno and Zurek}
\address[A]{Informatics Institute, University of Amsterdam, Amsterdam, the Netherlands}
\address[B]{Institute of Mathematics and Computer Science, Maria Curie-Sklodowska University, Lublin, Poland}

\begin{abstract}
Contemporary AI alignment and safety efforts treat system governance largely as an empirical engineering task, often without connection to fundamental research conducted on normative systems and ethical theory. In this paper, we propose a unifying conceptual architecture for technical governance informed by major ethical constructs. Rather than favoring one construct over the others, our work focuses on their core, complementary regulative functions. Namely, we model norms as satisfiability constraints and values as optimization attractors; we qualify deontic and consequentialist constructs through a control-system perspective, and approach virtue-ethics constructs from a compositional perspective. We then extend this mapping to meta-level constructs—looking at contractualism, discourse ethics, and care ethics—operating on the relations between individual and collective levels of agency. The resulting framework aims to offer a map to qualify given technical solutions, as well as to identify possible additional regulative interventions.\end{abstract}

\end{frontmatter}
\markboth{October 2026\hb}{October 2026\hb}

\section{Introduction}

The need to regulate the behaviour of artificial systems is today a matter of urgent societal relevance and motivates research within and across several disciplines. In computational domains, the matter becomes one of \textit{technical governance}, i.e. having effective instruments (i) for regulating (driving or evaluating) behaviour given certain normative dispositions, as well as (ii) for evaluating the normative dispositions themselves. These matters fall within traditional research tracks in AI \& Law, generally grouped under the umbrella terms of \textit{normative systems} and \textit{computational (theory of) law}: how normative directives can be (formally, and thus symbolically) specified to be machine-processable \cite{alchourron1984pragmatic,jones1992deontic,governatori2022thirty}, and how they can be applied in execution (for intervening in the selection of the system behaviour in a certain context) and in interpretation (e.g. for verifying whether behaviour of another actor is compliant, or for checking conformance with other normative dispositions)  \cite{breuker2005use,sartor2022thirty}. Yet, looking at contemporary AI alignment and AI safety research, remarkably little of that foundational effort has carried over. Most current approaches attempt to identify and modify values in AI (primarily LLM-centered systems)—either directly within sub-symbolic models through e.g. \textit{mechanistic interpretability} \cite{bereska2024mechanistic} and then activation-steering \cite{zhang2025} or just by \textit{fine-tuning} with adequate data, or by investigating and regulating the textual input/output couplings, by relying on natural-language artifacts to study expressed or revealed values (e.g., \textit{value extraction}, for instance from model responses to dilemmas \cite{chiu2025dailydilemmas}) and guide self-regulation (e.g. \textit{constitutional AI}) \cite{bai2022constitutional}. % In practice, these artefacts have the role of internal policy documents meant to govern a deployed AI system as would be an organizational actor. 
Yet, hybrid architectures exist, that attempt to make subsymbolic and symbolic components interact, typically employing symbolic deontic guardrails after a particular behaviour has been selected, or intervening in the behavioural selection process by a coordination of subsymbolic activation and symbolic derivation (e.g. \cite{chatterji2026plguard}). 

As this brief outline suggests, these contributions attempt to solve the behavioural regulation problem primarily as an engineering problem, rather than developing a more general theory of how technical governance should actually function. This broader aim would share motivations with early \textit{cybernetics}, although it differs slightly in its framing. Whereas that discipline focused primarily on control systems realized in biological systems, electronic devices, and models of organizations seen through organicist metaphors, here we frame the problem as one of \textit{distributed governance} (and thus relying on forms of \textit{decentralized control} applied by autonomous entities), as observed in human social systems as well as in complex computational infrastructures such as the Internet \cite{sileno2021accounting}. Normative multi-agent systems (NMAS) research \cite{alechina2013computational,balke2013norms}  originated from a similar distributed, open perspective, although practical computational objectives/constraints arguably limited the scope of representation of most contributions. One alternative way to approach this endeavour would be to make sense of human constructs concerning the regulation of behaviour at large, traditionally discussed within the branch of philosophy known as \textit{ethics}. A few works have taken this route (see e.g. \cite{Berreby2017ModularEthics,bartocci2021enforcing,wyner2024towards,zurek2024virtue}, or the literature study of a few years ago  \cite{tolmeijer2021}); yet, these works tend to focus on only a few ethical constructs (typically deontic vs consequentialist, treating collective norms as guardrails to individual values), rather than integrating them into a shared framework.   
Wider questions remain, such as \textit{how can deontological and axiological perspectives, deontic and consequentialist approaches, virtue ethics, contractualism, discourse and care ethics, as well as individual and collective levels of agency, be given a shared computational interpretation?}

The elaboration we propose in this paper does not aim to provide an accurate and complete account of these ethical traditions nor of the debates around them. Its purpose is primarily to explore how different ethical perspectives can be captured in modular, computationally meaningful terms; the proposed models are thus by design simplified interpretations. Yet, refraining from privileging one ethical perspective over another, and aiming to form a unified computational model of behavioural regulation, we can take the complementarity of the proposed definitions as initial evidence of \textit{representational adequacy}.

\section{Satisfiability of norms vs Optimization of values}

While evaluating behaviour, a classical distinction is drawn in philosophy between judgments of \textit{right/wrong} conduct and judgments of \textit{good/bad} conduct. Two areas of ethical inquiry are closely related to this distinction: the study of duties and normative requirements, central to \textit{deontology}, and the study of values, core to \textit{axiology}. % \footnote{For introductory overviews of deontological ethics, consequentialism, and value theory, see e.g. Larry Alexander and Michael Moore, ``Deontological Ethics,'' \url{https://plato.stanford.edu/entries/ethics-deontological/}; Walter Sinnott-Armstrong, ``Consequentialism,'' \url{https://plato.stanford.edu/entries/consequentialism/}; and Mark Schroeder, ``Value Theory,'' \url{https://plato.stanford.edu/entries/value-theory/}.} 
The former typically employs deontic vocabulary to describe conduct: \textit{obligations}, \textit{prohibitions}, and \textit{permissions} (and, with various nuances, \textit{liberties}), whereas the latter concerns the goodness or badness of actions, outcomes, states of affairs, or character traits. For modeling purposes, norm-based evaluation asks whether conduct abides by applicable requirements; value-based evaluation asks what that conduct promotes or demotes. These perspectives do not strictly correspond to a distinction between procedures and outcomes: norms may require particular outcomes, and values may attach to procedures themselves. Similarly, deontology does not hold a monopoly on defining what is right and wrong. One can adopt a value-based perspective to evaluate the rightness of actions---for instance, by defining right conduct as that which maximizes total value (as in reinforcement learning). Dually, one can introduce normative standards through (legal) \textit{principles} in order to promote or protect certain values (see e.g. \cite{araszkiewicz2025tipping}), and thus declaring what is deemed to be good or worth to strive for and protect.   

\paragraph{Norm compliance vs value promotion}
Yet, a structural distinction can still be drawn between norm compliance and value promotion. With respect to a concrete, applicable obligation or prohibition, given the information needed to assess it, conduct may either comply with or violate the requirement. Evaluation is thus modeled as binary. Conflicting requirements may be addressed through given priorities, overriding mechanisms, or, if these do not apply, procedures for determining which requirements prevail (generally asking for human intervention). Eventually, these mechanisms serve to resolve the conflict, so that externally, normativity appears as univocally directed, otherwise norm compliance evaluations may not have a binary conclusion. By contrast, conduct may promote or undermine values to different degrees, or leave them unaffected. Value-based evaluation is therefore naturally modeled through graded assessments or relative rankings. Such assessments need not presuppose a numerical scale: they may instead yield a partial ordering of alternative scenarios within and across several value dimensions. Where suitable numerical representations and aggregation rules are available, optimization can combine these assessments. Aggregation across several value dimensions can therefore support a choice among conflicting considerations (e.g. through weights) while retaining the trade-offs in the individual components. %, provided those components remain available. 

\paragraph{Satisfiability and optimization}
From a computational point of view, the observed duality suggests a correspondence between norm compliance and problems of \textit{satisfiability} or \textit{constraint satisfaction}, on the one hand, and between value promotion and problems of \textit{optimization}, on the other. The former ask whether there is an assignment or a course of action satisfying the specified constraints; the latter ask which alternative is best according to a specified objective and within a given feasible or acceptable set of conducts.\footnote{In engineering more widely, the corresponding distinction is between feasibility problems (\textit{is there a solution satisfying the requirements?}) and benchmarking/optimization problems (\textit{which feasible solution is best?}).} % Assessing whether a given course of action satisfies the requirements is a related but distinct task from establishing whether any satisfying course of action exists.
Evidently, although satisfiability and optimization are distinct tasks, they are connected. A constraint-satisfaction problem can, for example, be reformulated as minimizing constraint violations. Dually, expressing constraints as value thresholds introduces \textit{satisficing} \cite{simon1955behavioral} directly into the optimization process. The optimization problem itself may be approached through  checks at different objective thresholds. Such reformulations do not, however, make the representations the same for practical use; their suitability depends on the structure of the problem, the information that must be preserved, and the task to be performed. The computational distinction is therefore a distinction between modeling norms (requirements) as \textit{conditions to be fulfilled} and modeling values as \textit{objectives to be pursued}.% \footnote{Yet, it is fair to observe that in contexts involving several unresolved violations, an axiological perspective appears more adequate for balancing nuances across dimensions, not least because it provides finer granularity for deciding which dimensions to prioritize.}

\paragraph{Regions, attractive and repulsive forces}
When requirements are modeled as \textit{hard constraints}, they delimit an ``acceptable'' region within which behavioural objectives are pursued. Where an appropriate encoding is available, objectives can be represented through \textit{soft constraints} (positively, as attractive forces, or negatively, as repulsive forces) --- see e.g. \cite{Fratric2023}. Yet, the delimitation of the acceptable region may exclude high-value behaviours, which raises the problem of evaluating the constraints themselves. 
It may be that given norms do not promote the values that regulation is meant to promote. More generally, hard constraints or their formalization may be incomplete, misinterpreted, or no longer appropriate. Assessing the adequacy of norms (and thus of the acceptable region) is an additional activity that needs to be maintained for proper regulation. 

\vspace{-5pt}

\section{Regulative interpretation of open- and closed-loop control}

In control theory, systems are generally divided into \textit{open-loop} control systems, where action is performed without feedback from the controlled output, and \textit{closed-loop} ones, in which feedback informs the commands to the actuator. In a typical negative-feedback arrangement, the controller uses the difference between a given reference and the observed condition (also known as ``error'') to generate corrective commands. The reference and the observed condition may vary over time, independently of the control system. Returning now to ethical constructs, a \textit{consequentialist} perspective can be modeled through the evaluation of the outcomes of conduct, abstracting away from differences in performances insofar as they do not affect the consequences considered relevant. Where observed outcomes inform subsequent action, they realize a closed feedback loop. If those outcomes become observable only after performance, evaluation is intrinsically delayed---a property, in practically-relevant domains, which maps to increasing \textit{uncertainty}. We can use the term \textit{mediated intent} to describe a control objective whose achievement is assessed through these consequences, rather than the conduct itself (and thus suffers some degree of uncertainty).\footnote{Importantly, a consequentialist evaluation (mediated intent) does not require waiting for physical outcomes to manifest in the external world. An agent may evaluate consequences in anticipation through simulation (see e.g. feed-forward control or \textit{model predictive control} methods).} % (Here we assume performance may be simulated so as to consider foreseen consequences).  
In contrast, a \textit{deontic} perspective can be modeled as prescribing conduct without focusing on consequences. Its implementation may then use open-loop control, where behaviour is commanded without feedback, or a closed loop in which assessment goes in parallel with performance. Because there is no delay (or uncertainty), %the control loop regulates performance, not outcome. 
we use \textit{immediate intent} for the latter case, as performance of the prescribed conduct provides in itself the assessment signal. %For instance, we know whether we are walking while we are walking; actually we do it without needing to be aware of it. % We saw some monitoring was in place as soon as something stopped us 

\section{Signaling interpretation}

Under the previous interpretation, certain conducts (the ones we called immediate intents) require less monitoring effort than others (mediated intents) whose assessment depends on more information due to delayed, uncertain, or difficult-to-attribute outcomes. This suggests that deontic constructs (immediate intents) do not only circumscribe behaviour (at the \textit{generation} side), but can also serve signaling purposes, particularly where the prescribed conduct is readily observable and its interpretation is sufficiently shared within a social group. They can be utilized to speed up the evaluation of compliance to declared requirements and therefore, in social terms, for \textit{in-group/out-group} selection (at the \textit{verification/discrimination} side). 

Consequentialist directives (mediated intents), on the other hand, even when outcomes can be expressed in binary form, may require dealing with \textit{uncertainty}, which may be evaluated differently by members of a community. Yet, when there is some reasonably shared understanding of \textit{foreseeable} consequences, this can inform deontic statements that prescribe conduct expected to produce those consequences.

\vspace{-5pt}
\section{Behavioural niches, at individual and collective level}

For modeling purposes, we define an \textit{individual behavioral niche} as a circumscribed operational setting in which an agent's conduct develops through interactions with its environment. In individual learning, concrete experiential feedback---such as pain and pleasure signals (or lack thereof)---informs internal value assessments, which gradually solidify into individual-sourced regulatory patterns or habits. Similarly, evolutionary processes shape innate behavioral dispositions over longer time horizons. We employ \textit{individual-sourced norms} as an umbrella term for these diverse mechanisms (including habits, innate instincts, and learned dispositions): functionally, they represent behavioral patterns that regulate conduct relative to expected and preferred environmental states.

For an individual entering a collective (socially constructed) behavioural niche, the development order may be reversed. For the lower cognitive burden they put on individuals (section 3, 4), norms are the first available level of regulation, which may subsequently be explained in terms of the collective values they are meant to promote. The reference to collective values becomes particularly important for handling \textit{hard cases}, where judgment is needed again, allowing the evaluation of the ``acceptable'' region boundary (section 2).

The interaction between individual and collective norms occurring in an agent can then be modeled as an adaptation to collective requirements, including the development of expectations and habits. Yet, adaptation does not necessarily imply \textit{endorsement} (taken as \textit{volitional alignment}) nor \textit{following} (taken as \textit{behavioral alignment}), through which the individual agent aligns more with the collective one. Individual drivers---encompassing personal norms and values---besides avoidance, may also foster political participation, through which the collective level itself is contested, negotiated, or modified, so that the collective agency aligns more with the individual one.

\vspace{-5pt}
\section{Revisiting ethical constructs}

Within the conceptual framework we propose, \textit{deontic} directives are modeled as qualifications on conduct (as performance), providing a foundational ingredient of normative reasoning. \textit{Consequentialist} evaluations are modeled as assessments of conduct (as outcomes), providing a foundational ingredient of value-based reasoning. We situate both constructs within a single ecological niche of action (i.e. a delimited domain of activity). We can then interpret \textit{virtue ethics} through a multi-niche perspective, in which relatively stable dispositions and practical judgment support coordination across the competing demands of distinct behavioral niches. Key computational mechanisms include priorities among norms, and balancing/weighing across multi-objective value spaces. The primary modeling concept here is that of a \textit{composite role}, which formalizes the coordination of primitive roles, each bound to the normative requirements or evaluative objectives of its underlying niche. A successful composition enables the agent to act as a cohesive entity across complex behavioral domains. Additionally, roles can be characterized both positively and negatively, corresponding to behavioral dispositions evaluated as virtuous or vicious roles within a social setting.

\textit{Contractualism} is approached through the dynamic relation holding between regulating and regulated entities, independently of whether regulation targets performance or outcomes. Its structural function is to formalize the bridge between individual and collective agency: capturing mechanisms of understanding, acceptance (endorsement, following) or rejection of specific behavioral requirements. % For instance, rule-following can be modeled as contingent upon explicit volitional acceptance or structural endorsement.

Finally, relational ethics such as \textit{discourse ethics} and  \textit{care ethics} can be seen as prescriptive mechanisms governing how the perspectives of agents affected by regulation are articulated and examined, and possibly acted upon. %In our framework, its function is to connect high-level rule evaluation directly to individuals' grounded material circumstances (e.g., experiential pain and pleasure), creating structured channels for political participation, feedback, and normative challenge. Particular weight is given to agents bearing disproportionate adverse impacts. 
Since in this paper we focused on the conceptual layer of the model, discussions on procedural aspects are left to future work. 
The key insight here is that, by applying this unified framework, these ethical perspectives emerge not as competing doctrines, but as complementary architectural components feeding an organic, yet decentralized, regulative system.

\vspace{-5pt}
\section{Sketch of Formalization}

We introduce a sketch of the formalization of the above conceptual framework. Note that this is not a fully fledged model, but rather an introduction of the general concepts and their relations. We purposely avoid discussing the details of specific deontic and value-based approaches (as well as the numerous variations of the analysed ethical theories) in order to keep the model sufficiently flexible. % and to focus on the relations between different approaches to ethics (deontology, consequentialism, and virtue) and between individual and social components. % Thanks to this, the model of open- and closed-loop control is left for future work, in which a consistent and holistic model will be presented. 

\subsection{Ground level}  

Formally, let an agent operate over a state space $\mathcal{S}$ and an action space $\mathcal{A}$. We consider two primitives of behavioural regulation: \textit{deontic primitives}, which specify whether an action $a$ is obligatory, permitted, or prohibited, and \textit{consequentialist primitives}, which specify how an action $a$ promotes or demotes a value $v$ with a certain intensity (possibly incorporating a measure of uncertainty). 

\paragraph{Deontic primitives} Let $\mathcal{N} = \{n_1, \dots, n_k\}$ be a set of deontic directives. As observed in section~5, these need not be social or legal norms, as they may also be individual-sourced norms; here, they should be interpreted as hard constraints to which the agent is committed. A primitive directive $n_i \in \mathcal{N}$ is a partial function $n_i: \mathcal{S} \times \mathcal{A} \rightharpoonup \{\mathrm{forb}, \mathrm{perm}, \mathrm{obl}\}$. The point-wise mapping can also be expressed in predicate form, e.g., $\mathrm{perm}_i(s, a)$. On top of these primitives, additional directives introducing priority mechanisms can be considered, including meta-rules concerning applicability, which serve to \textit{procedurally} resolve potential conflicts emerging during derivation.
Based on the three modalities, a set of directives $\mathcal{N}$ defines three regions in $\mathcal{S} \times \mathcal{A}$. For instance, the permission region $\mathcal{P} \subseteq \mathcal{S} \times \mathcal{A}$ of conduct is defined as:
\begin{equation}
\mathcal{P} = \{ (s, a) \in \mathcal{S} \times \mathcal{A} \mid \mathcal{N} \vdash \mathrm{perm}(s, a) \}
\end{equation}

We can similarly define the forbidden region ($\mathcal{F} \subseteq \mathcal{S} \times \mathcal{A}$) and the obligatory region ($\mathcal{O} \subseteq \mathcal{S} \times \mathcal{A}$).
The \textit{acceptable region} $\mathcal{AR}$ can then be defined on the basis of $\mathcal{P}$, $\mathcal{O}$, and $\mathcal{F}$, depending on the chosen deontic axiomatization and normative closure defaults. For instance, in a multi-agent setting, interpreting permissions as protected liberties maps them to prohibitions against interfering, which are added to $\mathcal{F}$. The acceptable region can then be defined, for example, as $\mathcal{AR} = \overline{\mathcal{F}}$ under a permissive default (everything not forbidden is acceptable), or as $\mathcal{AR} = \mathcal{P} \cup \mathcal{O}$ under a restrictive default (only explicitly permitted or obligatory actions are acceptable).

Regardless of this choice, norm compliance maps to a constraint-satisfaction problem: evaluating whether a conduct $(s, a)$ belongs to the acceptable region $\mathcal{AR}$ yields a binary verdict. We denote by $\mathcal{AR}(s)$ the set of actions acceptable in $s$. If no action is acceptable in a given state $s$ ($\mathcal{AR}(s) = \varnothing$), an unresolved normative conflict has emerged.

\paragraph{Consequentialist primitives} 
Conversely, value promotion is modeled via a value function $v: \mathcal{S} \times \mathcal{A} \to \mathbb{R}$, which captures performance along a specific axiological dimension. At a primitive level, value-based optimization seeks an action choice $a^*(s)$ that maximizes $v$ within the acceptable region:
\begin{equation}
a^*(s) = \arg\max_{a \in \mathcal{A}} \; v(s, a) \quad \text{subject to } (s, a) \in \mathcal{AR}
\end{equation}
A purely consequentialist agent has an acceptable region $\mathcal{AR} = \mathcal{S} \times \mathcal{A}$ (no hard constraints are in place); a purely deontic agent has $v(s, a) = 0$ for all $(s, a) \in \mathcal{S} \times \mathcal{A}$, or no value dimension to maximize (no optimization objective is in place), so, in $s$, executing any action in $\mathcal{AR}(s)$ results in compliant behaviour. Note that a consequentialist-defined agent can be (lossily) recast as a deontic one: under \textit{strong selection} (maximizing), it can be governed by the obligation to select the optimal action $a^*(s)$ in each state $s$; dually, under \textit{weak selection} (satisficing), it can be governed by the prohibition against selecting actions whose value falls below a minimal threshold $t$. Operationally, this mapping clarifies the distinction between \textit{immediate} and \textit{mediated intent} perspectives (section~3). Under mediated intent, the agent retains a space of action governed by the expectations encoded in $v(s, a)$, which can be updated in real time as soon as feedback is observed from the environment. Under immediate intent, no such space is granted: conduct is assessed directly against $\mathcal{AR}$, and any action outside it is immediately flagged as a violation. Because law generally aims to \textit{circumscribe} the behaviour of autonomous agents, its deontic dimension naturally becomes more prominent. In other words, law compiles out minimal thresholds of value satisficing expected to be maintained by social participants.

\subsection{Multi-niche perspective}  

A central modeling insight of our architecture concerns where aggregation (when multiple value dimensions are at stake) and conflict resolution (when directives are incompatible) take place. We place these mechanisms explicitly in a role (more precisely, \textit{agent-role}) layer.

\paragraph{Balancing} 

Suppose an agent is guided by a set of distinct values, captured through a multi-dimensional value function $\mathcal{V} = (v_1, \dots, v_n)$. To balance, compare, or weigh different value dimensions during optimization, the agent must possess or construct an \textit{aggregation} function over the vector of value functions, yielding a composite value function $v^* = \Gamma(\mathcal{V})$ with $v^*: \mathcal{S} \times \mathcal{A} \to \mathbb{R}$ (we do not commit to a specific functional form; while a weighted sum is a common choice, non-linear aggregations are also possible). Balanced value-based optimization then seeks an action $a^*(s)$ that maximizes this aggregated value within the acceptable region:
\begin{equation}
a^*(s) = \arg\max_{a \in \mathcal{A}} \; v^*(s, a) \quad \text{subject to } (s, a) \in \mathcal{AR}
\end{equation}
where $\mathcal{AR}$ denotes the acceptable region entailed by the deontic component.

\paragraph{Conflict resolution} 

Suppose now that the agent is addressed by a set of distinct normative sources $\mathfrak{N} = \{\mathcal{N}_1, \dots, \mathcal{N}_m\}$. Taking the combined normative system as the simple union $\mathcal{N} = \bigcup_{i=1}^m \mathcal{N}_i$ may give rise to normative conflicts. In a geometric interpretation, a single conflict can be equivalently modeled as an empty intersection of the  acceptable regions determined by the various sources at a given state $s$, i.e., $\bigcap_{i=1}^m \mathcal{AR}_i(s) = \varnothing$. In such cases, the agent must decide which norms to relax or sacrifice in order to retain an actionable behavioural directive (noting that inaction itself counts as a behavioural decision). Even without specifying the exact resolution procedure, value dimensions and their balancing clearly play a role in this decision; moreover, because conflicts emerge within specific situations, their resolution should likewise be defined at a state-level rather than in a purely abstract sense:
\begin{equation}
\theta: \mathfrak{N} \times \mathcal{V} \times \mathcal{S} \rightarrow \Delta \mathcal{N}
\end{equation}
Here, the output $\Delta \mathcal{N}_s = \theta(\mathfrak{N}, \mathcal{V}, s)$ denotes a set of meta-rules encoding overriding mechanisms for the norms  conflicting in $s$, such that $\mathcal{N} \cup \Delta \mathcal{N}_s$ yields a non-empty acceptable region $\mathcal{AR}'(s) \neq \varnothing$. Note that because balancing informs conflict resolution, $\theta$ is generally parameterized by—or functionally dependent on—the aggregation mechanism $\Gamma$ (so, it should rather be denoted as $\theta_\Gamma$). This schematization connects directly with case-based reasoning in AI \& Law, where precedential rules are distinguished or overridden on the basis of underlying value considerations \cite{benchcapon2003persuasion,prakken2015formal}.  Operationalizing how values guide such overrides typically involves value preference orderings, optimization, or thresholding mechanisms---for instance, triggering a norm override only when the resulting value promotion or harm mitigation exceeds a critical threshold, or when value levels fall below minimal acceptability bounds \cite{araszkiewicz2025tipping,zurek2024virtue}. From an operational point of view, given $\mathcal{N}$ and $\theta_\Gamma$, we can define a \textit{consolidation} function that returns a revised rule base in which rules are rewritten exclusively in terms of the deontic primitives resulting from conflict resolution (cf. \textit{priority-based} vs \textit{constraint-based} representations in \cite{Sileno2015}).

\paragraph{Primitive and composite agent-roles}

Let us consider a set of behavioural niches $\mathcal{K}$ (e.g., activities related to social roles such as being a parent, a teacher, a partner, a researcher, and so on). Each primitive niche $k \in \mathcal{K}$ introduces its own set of deontic directives $\mathcal{N}_k$, alongside a vector of value functions $\mathcal{V}_k$.

We model a \textit{primitive role} as a tuple $\rho_k = (\mathcal{N}_k, \Gamma_k, \theta_k)$, where $\mathcal{N}_k$ is the set of deontic directives associated with niche $k$, $\Gamma_k$ is an aggregation function over the niche's value dimensions $\mathcal{V}_k$, and $\theta_k$ is a local conflict-resolution function for role $k$. If resolution and aggregation succeed, an agent enacting a primitive role $\rho_k$ obtains a well-defined optimization objective $v^*_k$ and a resolved acceptable region $\mathcal{AR}_k$ to guide its behaviour. Therefore, for regulation to be functional, a behavioural niche must map to at least one well-defined primitive role.

A \textit{composite role} $R = \bigotimes_{k \in \mathcal{K}'} \rho_k$ (where $\mathcal{K}' \subseteq \mathcal{K}$) coordinates behaviour across multiple niches. In this schema, ``practical wisdom'' (\textit{phronesis}) acts as a higher-order policy represented by a function $\pi_R$ that aggregates the local aggregation functions $\Gamma_k$ across the constituent roles, and consequently, resolves possible normative conflicts. Specifically, if $\Gamma_{\mathcal{K}'} = \{\Gamma_k \mid k \in \mathcal{K}'\}$ is the set of aggregation functions associated with the active niches $\mathcal{K}'$, then $\pi_R(\Gamma_{\mathcal{K}'}) = \Gamma_R$. This meta-aggregation enables the resolution of cross-niche value trade-offs and provides the necessary axiological basis to arbitrate competing normative directives across niches via a composite conflict-resolution function $\theta_{\Gamma_R}$. Denoting by $\mathcal{N}_R = \bigcup_{k \in \mathcal{K}'} \mathcal{N}_k$ the union of all normative directives across $\mathcal{K}'$, the composite role can thus be expressed in the same canonical tuple form as $R = (\mathcal{N}_R, \Gamma_R, \theta_{\Gamma_R})$. Behavioural regulation then reduces to the unified constrained-optimization problem:
\begin{equation}
a^*(s) \in \arg\max_{a \in \mathcal{A}} \; v^*_R(s, a) \quad \text{subject to } (s, a) \in \mathcal{AR}_R
\end{equation}
This equation captures single-niche primitive roles ($|\mathcal{K}'| = 1$), purely consequentialist ($\mathcal{AR}_R = \mathcal{S} \times \mathcal{A}$), and purely deontic ($v^*_R \equiv 0$) as special cases.

\subsection{Relations between individual agent and collective agency}

Finally, we model how an individual agent's deontological (norms or acceptable regions) and axiological (values or driving forces) positioning relates to a collective agency. We can distinguish two operational attitudes: (1) the individual agent may be practically (deontologically) aligned with the collective agency, performing a behavioural selection that \textit{externally} satisfies compliance; (2) the individual agent may be volitionally (axiologically) aligned with the collective agency, when the collective directives are \textit{internally} compatible with the agent's value boundaries. This distinction enables us to introduce further concepts, such as norm-following and endorsement of given social norms.

\paragraph{Norm-following}
Illustratively, we can say that for an individual agent $i$ with normative constraints $\mathcal{N}_i$, a set of social norms $\mathcal{N}_c$ associated to a collective agency $c$ is:
\begin{itemize}
    \item \textit{strongly acceptable} if $\mathcal{AR}_i = \mathcal{AR}_c \wedge \mathcal{O}_i = \mathcal{O}_c \wedge \mathcal{F}_i = \mathcal{F}_c$;
    \item \textit{weakly acceptable} if not strongly acceptable, and $\mathcal{AR}_i \cap \mathcal{F}_c = \varnothing \wedge \mathcal{O}_c \cap \mathcal{F}_i = \varnothing$;
    \item \textit{weakly unacceptable} if $(\mathcal{AR}_i \cap \mathcal{F}_c \neq \varnothing \vee \mathcal{O}_c \cap \mathcal{F}_i \neq \varnothing)$ and $\mathcal{O}_i \cap \mathcal{F}_c = \varnothing$;
    \item \textit{strongly unacceptable} if $\mathcal{O}_i \cap \mathcal{F}_c \neq \varnothing$.
\end{itemize}
This classification defines a function qualifying norm-following: $\Delta_i: \mathfrak{N} \times \mathfrak{N} \to \mathcal{L}$, where $\mathcal{L}$ denotes the ordered set of acceptability labels above.

\paragraph{Endorsement of norms}
Values can serve as a basis for evaluating norms, meaning that an individual agent can endorse norms to varying degrees. When evaluating deontic directives (such as moral and legal norms), agents consider the outcomes that these directives oblige, permit, or prohibit achieving.
Illustratively, suppose a collective agency with norms $\mathcal{N}_c$ for which $(s,a) \in \mathcal{O}_c$, and a vector $T^i$ of minimal acceptable thresholds across the value dimensions $\mathcal{V}$ of agent $i$:
\begin{itemize}
    \item if $\forall v \in \mathcal{V}: v(s,a) \succeq T^i_v$, then the obligation of $a$ in $s$ is \textit{strongly acceptable};
    \item if $\neg \exists v \in \mathcal{V}: v(s,a) \succeq T^i_v$, then the obligation of $a$ in $s$ is \textit{strongly unacceptable};
    \item otherwise, it is \textit{weakly acceptable}.
\end{itemize}
Analogous pointwise conditions can be defined for $\mathcal{F}_c$ and $\mathcal{P}_c$. Aggregating these assessments across the directives of $\mathcal{N}_c$ would yield a function qualifying the \textit{endorsement} of social norms by agent $i$: $\Pi_i: \mathfrak{N} \times \mathbb{R}^{|\mathcal{V}|} \to \mathcal{L}$. Complete acceptance can then be defined on the basis of both following ($\Delta_i$) and endorsing ($\Pi_i$) acceptances.

\paragraph{Contractualist perspectives}
The illustrative formalization above is informed by a contractualist stance, in the sense that it focuses on the relations holding between an individual agent and a collective agency of reference. Interestingly, it can be applied both as a \textit{prescriptive} lens on norm adoption (in the context of a formal, top-down governance via contract-like artifacts) and as a \textit{descriptive} lens (in the context of informal, bottom-up conventions, including social-contract perspectives), which emerge practically as agents adapt to local community practices over time.

\paragraph{Relational interventions}
Building on these relations, \textit{discourse ethics} procedurally requires (i.e. via collective norms) that the collective agency take into account the perspectives of individuals who suffer adverse consequences from introducing or applying a given policy. This procedural attempt at \textit{volitional alignment} of the collective toward the individual is intended, when successful, to foster individual endorsement, as majority mechanisms tend to suppress minority or individual voices. However, volitional alignment alone may be insufficient to rebalance power asymmetries among social participants. At a functional level, \textit{care ethics} complements discourse ethics by requiring that once individual suffering or vulnerability is recognized (substantively, in terms of values), specific protective or assistive mechanisms be instituted at the collective normative level to repair this imbalance. The formalization of these constructs is left to future work.

\section{Conclusion}
The paper introduces an integrated framework attempting to unify different strands of discourse in ethics and technical regulation. The exercise has been fruitful, in that the theoretical pieces unexpectedly fit into complementary roles within a shared architecture. We must, however, reiterate that we do not claim to resolve the foundational normative debates among the ethical traditions addressed—in fact, there remains substantial value in keeping those philosophical discussions alive. Rather, the principal contribution of our work lies in isolating the specific functional roles each theory plays within an integrated control loop involving both individual agents and collective entities. Whereas existing computational approaches to AI ethics typically focus on modeling and implementing isolated theories \cite{wyner2024towards,zurek2024virtue,bartocci2021enforcing,tolmeijer2021}, we propose a different approach: by relying on minimal, baseline representations of diverse ethical theories we can examine their dynamic interactions at both the individual and collective levels.

\bibliographystyle{vancouver.bst}
\bibliography{references}
\end{document}